\documentclass[authoryear,5p, preprint]{elsarticle}

\usepackage[utf8]{inputenc} 
\usepackage[T1]{fontenc}    
\usepackage{hyperref}       
\usepackage{url}            
\usepackage{booktabs}       
\usepackage{amsfonts}       
\usepackage{nicefrac}       
\usepackage{microtype}      
\usepackage{fullpage}
\usepackage{lineno}
\usepackage{natbib}
\usepackage{pifont}
\usepackage{amsmath,amssymb}
\usepackage{graphicx}
\usepackage{subcaption}
\usepackage{multicol,graphicx,xcolor,multirow, array}
\usepackage[normalem]{ulem}
\usepackage{mathtools}
\usepackage{accents} 
\usepackage[export]{adjustbox} 
\usepackage{diagbox}

\usepackage{tikz}
\usetikzlibrary{shapes.misc, positioning}
\usetikzlibrary{matrix, calc, positioning, arrows,shapes,backgrounds, arrows.meta, quotes}
\usepackage{gensymb,stmaryrd,mathtools,textcomp,xspace,grffile}

\usetikzlibrary{shapes.geometric,fit,matrix,positioning,shapes.multipart}
\usetikzlibrary{decorations.markings}
\usetikzlibrary{shadows,decorations.pathreplacing,fadings}
\pgfdeclarelayer{background}
\pgfsetlayers{background,main}
\usetikzlibrary{topaths, calc,3d}
\usetikzlibrary{fit}

\cortext[cor1]{Corresponding author}

\begin{document}

\begin{frontmatter}

\title{DeepMerge: Classifying High-redshift Merging Galaxies with Deep Neural Networks}

\author[matf,sanu,fnal]{A.~\'Ciprijanovi\'c\corref{cor1}}
\ead{aleksandra@matf.bg.ac.rs}
\author[stsci]{G.F.~Snyder}
\author[fnal,kicp,uofcaa]{B.~Nord}
\author[stsci,jhu]{J.E.G.~Peek}

\address[matf]{Department of Astronomy, Faculty of Mathematics, University of Belgrade, Studentski trg 16, 11000 Belgrade, Serbia}
\address[sanu]{Mathematical Institute of the Serbian Academy of Sciences and Arts, Kneza Mihaila 36, 11000 Belgrade, Serbia}
\address[fnal]{Fermi National Accelerator Laboratory, P.O. Box 500, Batavia, IL 60510, USA}
\address[kicp]{Kavli Institute for Cosmological Physics, University of Chicago, Chicago, IL 60637, USA}
\address[uofcaa]{Department of Astronomy and Astrophysics, University of Chicago,IL 60637, USA}
\address[stsci]{Space Telescope Science Institute, 3700 San Martin Drive, Baltimore, MD 21218, USA}
\address[jhu]{Department of Physics \& Astronomy, Johns Hopkins University, 3400 N. Charles Street, Baltimore, MD 21218, USA}

\begin{abstract}

We investigate and demonstrate the use of convolutional neural networks (CNNs) for the task of distinguishing between merging and non-merging galaxies in simulated images, and for the first time at high redshifts (i.e.,  $z=2$).
We extract images of merging and non-merging galaxies from the Illustris-1 cosmological simulation and apply observational and experimental noise that mimics that from the Hubble Space Telescope; the data without noise form a ``pristine’’ data set and that with noise form a ``noisy’’ data set.
The test set classification accuracy of the CNN is $79\%$ for pristine and $76\%$ for noisy. 
The CNN outperforms a Random Forest classifier, which was shown to be superior to conventional one- or two-dimensional statistical methods (Concentration, Asymmetry, the Gini, $M_{20}$ statistics etc.), which are commonly used when classifying merging galaxies.
We also investigate the selection effects of the classifier with respect to merger state and star formation rate, finding no bias.
Finally, we extract Grad-CAMs (Gradient-weighted Class Activation Mapping) from the results to further assess and interrogate the fidelity of the classification model. 

\end{abstract}


\end{frontmatter}

\section{Introduction}
\label{sec:introduction}

Galaxy mergers are a primary trigger and probe of the evolution of cosmic structures.
The hierarchical merging of galaxies is both a probe of the cosmos as a whole to test the canonical $\Lambda$CDM cosmology paradigm  ~\citep{TT1972,KW1993,GW2008,C2014,RG2017} and a laboratory for the evolution of galaxies as astrophysical objects ~\citep{RO1977,WR1978}.
A particularly interesting period is "cosmic high noon," which took place at redshifts $z\sim2-3$. 
During this period, star formation rates are the highest, and significant amounts of stellar mass are assembled into galaxy-scale bodies~\citep{MD2014}. 
In the context of galaxy mergers, this period is still not fully understood. 
Several recent empirical studies have discovered evidence that the rate of occurrence of major merging events may become constant or start to decrease during the period $1< z <3$~\citep{RC2008,MZ2016}. 
This disagrees with theoretical models~\citep{HB2010,RG2015}, which predict that major merger rates continue to rise during this period.
Counting merger rates during this cosmic epoch may aid in or lead to explanations for the appearance of galaxies today, and shed light on the importance of mergers in galaxy evolution. 

Detecting galaxy mergers in observations by conventional automated methods or by visual inspection has proven to be quite expensive and time-consuming ~\citep{PP2002,LK2004,BJ2000,LS2011}.
One method of detection is selecting close galaxy pairs --- visually in the plane of the sky and in redshift ~\citep{BG2000,LK2004}. 
This method depends on the availability of deep, broadband multi-wavelength or spectroscopic data. 
These methods also suffer from the inability to distinguish between close pairs of galaxies that will eventually merge and those that will just pass by each other, resulting in sample contamination by galaxy flyby's~\citep{PB2013,LH2014,KP2014}.
Searching for merging pairs of galaxies can be done by visual inspection by large numbers of people --- e.g., GalaxyZoo;~\cite{LS2011}.
However, this process will become prohibitively time-consuming as data volumes increase and is subject to the biases of human classifiers.
High-resolution and high signal-to-noise images are required when merger classification is performed with parametric measurements of structure.
Examples include the S\'ersic index~\citep{S1963}, the Gini coefficient, the second-order moment of the brightest $20\%$ percent of the galaxy’s flux $\mathrm{M}_{20}$~\citep{LP2004}, {\it CAS} -- Concentration, Asymmetry, Clumpiness ~\citep{CB2003}, and identification of concentrated galaxy nuclei at small separations identified through median-filtering ~\citep{Lackner2014}.
The need for high-quality observations means that space-based observations are the only way to perform morphological analysis at higher redshifts ($z>1$). 
Small samples of observed distant galaxies introduce uncertainties in the study of galaxy merger history (this will improve with future missions like WFIRST\footnote{https://wfirst.gsfc.nasa.gov/index.html}, which will provide large volumes of data).

In recent years, classification tasks and learning from large data sets are often performed using neural networks -- a type of model for learning algorithms comprised of computational neurons, each of which has adjustable parameters (a weight and a bias). 
These parameters are adapted under the response to discrepancies between a network prediction and a truth label. 
The loss encodes the discrepancy, and this discrepancy is used to update the weights for each neuron using backpropagation: this procedure calculates the gradient of the loss function with respect to the neural network's weights -- a typical method for this is Stochastic Gradient Descent ~\citep{KW1952}.

Convolutional Neural Network (CNNs) are a primary representative of deep learning algorithms. They can be optimized for computer vision tasks, which makes them a good tool to use with astronomical images.
An important advantage of CNNs is in their capacity to discern patterns in large and complex data sets. 
These algorithms also do not require parametrically defined prior information about physical parameters (i.e., features) of the objects that are subject to measurement or classification. However, CNNs learn from a training set that has labels for ``ground truth'', and prior information enters in this form.
In cases where real observations do not offer large enough labeled image data sets (or when it is difficult to label observed images with enough certainty), training CNNs can be performed using images from simulations, which can often be made to be very large and diverse. 

CNNs have already proved very useful across a broad range of astronomical tasks --- e.g., identification of strong lensing events ~\citep{PT2019,JC2019}, lensing reconstruction of the Cosmic Microwave Background~\citep{CW2018}, identification of distant galaxies in a central blue nugget phase~\citep{HC2018}, learning galaxy morphology~\citep{DS2019}, identification of low-surface brightness tidal features in galaxies~\citep{WF2019}, classification of the large-scale structure of the universe~\citep{AC2019}, learning parameters that describe the first galaxies from 21-cm tomography of the cosmic dawn and reionization~\citep{GM2019} etc. 

CNNs have been used in a few cases in the context of galaxy mergers --- classifying at low-redshift ~\citep{AS2018,PWA2019,PW2019}, prediction of merger stage~\citep{BH2019}.
CNN performance depends on the type of training images, and training on galaxies extracted from large-scale simulations can be successfully used for detecting merging galaxies in real survey data ~\citep{AS2018, PW2019}. These strategies have not yet been applied to high-redshift galaxies.

The remainder of the paper is organized as follows. In \S\ref{sec:data}, we present the simulated data sets with which we train and test our algorithm, and in \S\ref{sec:deepmerge} we describe the implementation of CNNs for classification. 
We then describe the results of classification of mergers by CNNs, including a comparison with the results from random forest implementations from other works in \S\ref{sec:results}. We discuss our results in \S\ref{sec:discussion}. 
Finally, we summarize, conclude, and present an outlook for future work in \S\ref{sec:conclusion}.

\section{Data}
\label{sec:data}
It is extremely difficult to obtain real-sky observational data of labeled mergers at high redshifts and in quantities that are typically sufficient for training supervised machine learning algorithms.
Therefore, simulated data is critical for this task. 
We use simulated data from the Illustris-1 cosmological simulations ~\citep{VG2014a,VG2014b} as the baseline data set to which we add observational effects like point spread function (PSF) and random sky shot noise to produce the images we use.

\subsection{Data: ``Pristine'' and ``Noisy'' Simulations}
\label{sec:data:simulations} 
 
It tends to be very slow to find and label enough real observational images to build a sufficiently large training sample for even the shallowest of effective deep neural networks.
In these situations, simulated images that mimic real observations can provide additional useful training samples.
Simulations also offer the opportunity to craft training sets from three-dimensional ``ground truth,'' which may circumvent some biases that would be caused by using a training set defined purely from curated two-dimensional observations.
The predictive performance on real-sky observations of a CNN classifier trained on simulated data (and later used on real observations) will strongly depend on how successfully the simulated images mimic real observations.
 
We follow~\citep{SR2018}, who use images of galaxy mergers from the Illustris-1 cosmological simulation, using snapshots made in $12$ time-steps over $0.5 < z < 5$.
We use the subset of $z=2$ galaxy images.
Objects in extracted images are classified as mergers if the merging event occurs during the $500\,\mathrm{Myr}$ window around the time the snapshot from the Illustris simulation was taken. 
Merging events of a given stellar mass ratio are defined from the merger trees computed by \citet{RG2015}:
the time window for designation as a merger was chosen to be long enough to capture signatures during a wide range of merger stages \citep{Lotz2008} --- enabling identification of subtler and slower mergers, but short enough to omit galaxies whose morphology is unaffected by merging. 
In the current work, we consider mergers with a stellar mass ratio of 0.1 or greater.

Merging objects are considered to be the positive class (``P'') and non-merging objects the negative class (``N'').
No matter which time window is chosen, any classification algorithm is likely to give some false positives (non-mergers which look like mergers) and false negatives (mergers that look like non-mergers), for the time windows and merger event durations that don't match. 
For example, a pair of galaxies could approach very slowly so that the merger event happens outside the chosen time window, or the merger event could happen so quickly that a merger shows no physical effects only a short time later.
An example is shown in~\cite{SR2018}.
Clumpy star formation is also likely to present false-positives.
Mock images in various broadband wavelength filters were generated by \citet{Torrey2015}. 
In this work, we use two HST wavelength filters -- ACS F814W (red) and WFC3 F160W (near-infrared) that show features in a wide range of redshifts ($z\approx1-3$). 
For objects at $z=2$, these filters probe near-UV ($\approx0.27$ microns), which reveals bluer features in galaxies, like star formation, clumps, and asymmetries. 
The visible blue/green light ($\approx0.5$ microns) in the rest frame shows redder features that tend to reveal stellar mass and mergers.
These two filters are also relevant to data from the CANDELS survey~\citep{KF2011,GK2011}, which has uniform, deep coverage in all fields for both filters.
This forms the baseline data set without observational effects of photon noise or the telescope point spread function.
In~\cite{SR2018}, the authors modify the images to reflect the observational qualities of the Hubble Space Telescope (HST) and James Webb Space Telescope (JWST). First, the baseline images were convolved with a model point-spread function (PSF) appropriate for each filter (our "pristine" data set).
Then, random sky shot noise (approximated by a normal distribution) was added to each pixel, such that the final noisy images achieve a $5\sigma$ limiting surface brightness of $25$ magnitudes per square arc-second (our "noisy" data set) -- labeled ``SB25'' (while their PSF-only dataset is labeled ``SB00''). 

\subsection{Data preparation}
\label{sec:datapreparation}

We prepare the simulated data to be used for training, validation, and testing in the CNN optimization and analysis. The snapshot $z=2$ which we use, contains images of $2233$ different galaxies. Galaxy images were made using four ``camera'' perspectives, which were used as independent objects in order to augment the number of galaxy images. Finally, the image sample we used with our CNN includes $8930$ images, each in two HST filters -- ACS F814W and WFC3 F160W (2 images were discarded because they lacked all needed filters). 
The sample is unbalanced with a ratio of $1624:7306$ mergers to non-mergers.
We apply additional data augmentation (horizontal and vertical flips, rotations by $90\, {\rm deg.}$ and $180\, {\rm deg.}$) to the mergers in the data set to produce a more balanced sample consisting of $8120$ mergers and $7306$ non-merger. 
There are images from the $z=2$ snapshot that were not used in Random Forest classification by~\cite{SR2018}, due to the very low signal-to-noise ratio in each pixel or pathological Petrosian radius measurements (these images have merger probability $P_{RF}=\mathrm{None}$ in Table 2 of~\cite{SR2018}). 
We nevertheless include these low-quality systems, because they will be present in real observational data, especially in case of high redshifts. 
We resized all pictures to $75\times75$ pixels and use two HST filters (in both pristine and noisy case), making our input to CNN have dimension of $2\times75\times75$ (we use "channel first" image data format). 
Before training our CNN, we divide our images into training, validation and testing sample ($70\%:10\%:20\%$).

All of the images used in this paper are available online. Original baseline images can be found on the Illustris web page\footnote{http://www.illustris-project.org/data/}. All resized images that we used (both pristine and noisy) are available as a MAST High Level Science Product -- DOI:10.17909/t9-vqk6-pc80\footnote{https://doi.org/10.17909/t9-vqk6-pc80}.

\begin{figure*}
   \centering
  \includegraphics[width=\linewidth]{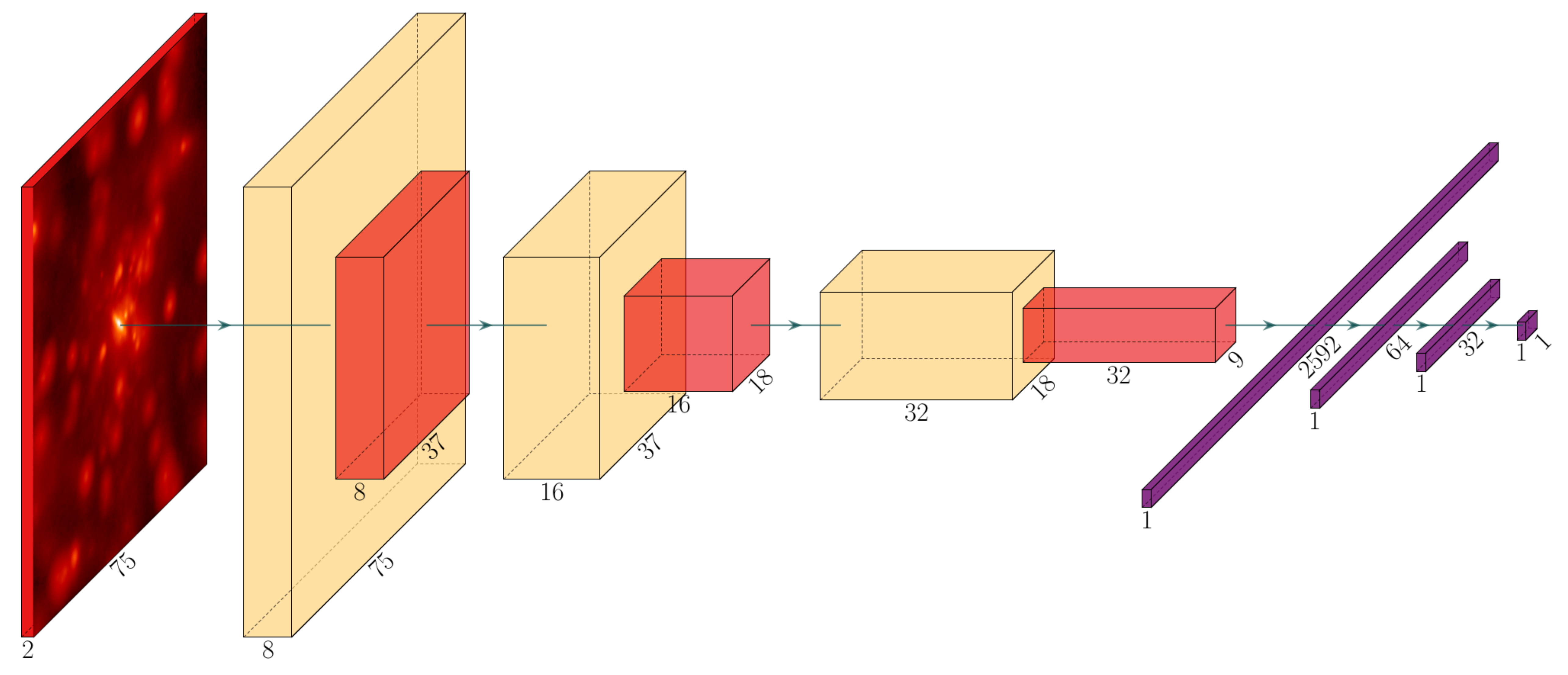}
\caption{
Architecture of the {\it DeepMerge} CNN presented in graphical form. 
Convolutional layers (three) are presented in yellow, pooling layers (three) in red, and dense layers (four - one after flattening and three additional that we add) in violet. 
Dropout layers are not shown.} \label{fig:arch}
\end{figure*}

\begin{table*}[!ht]
  \centering
  \noindent\begin{minipage}[b]{0.99\textwidth}
   \centering
    \caption{Architecture of the {\it DeepMerge} CNN.}
  \label{table:arch}
  \centering
  \begin{tabular}{|l | l l l  l   l |}
\hline
\bf{Layers}         & \bf{Properties}           & \bf{Stride}       & \bf{Padding}  & \bf{Output Shape} & \bf{Parameters}   \\ \hline\hline
Input               & $2\times75\times75$\footnote{We use "channel first" image data format.}       & -                 & -             & (2, 75, 75)       & 0                 \\ \specialrule{.2em}{.1em}{.1em}
Convolution (2D)    & Filters: 8                & $1\times1$        & Same          & (8, 75, 75)       & 408               \\
                    & Kernel: $ 5\times5$       & -                 & -             & -                 & -                 \\
                    & Activation: ReLU          & -                 & -             & -                 & -                 \\ \hline
Batch Normalization & -                         & -                 & -             & (8, 75, 75)       & 300               \\ \hline
MaxPooling          & Kernel: $2\times2$        & $2\times2$        & Valid         & (8, 37, 37)       & 0                 \\ \hline
Dropout             & Rate: $0.5 $              & -                 & -             & (8, 37, 37)       & 0                 \\ \specialrule{.2em}{.1em}{.1em}
Convolution (2D)    & Filters: 16               & $1\times1$        & Same          & (16, 37, 37)      & 1168              \\
                    & Kernel: $ 3\times3$       & -                 & -             & -                 & -                 \\
                    & Activation: ReLU          & -                 & -             & -                 & -                 \\ \hline
Batch Normalization & -                         & -                 & -             & (16, 37, 37)      & 148               \\ \hline
MaxPooling          & Kernel: $2\times2$        & $2\times2$        & Valid         & (16, 18, 18)      & 0                 \\ \hline
Dropout             & Rate: $0.5 $              & -                 & -             & (16, 18, 18)      & 0                 \\ \specialrule{.2em}{.1em}{.1em}
Convolution (2D)    & Filters: 32               & $1\times1$        & Same          & (32, 18, 18)      & 4640              \\
                    & Kernel: $ 3\times3$       & -                 & -             & -                 & -                 \\
                    & Activation: ReLU          & -                 & -             & -                 & -                 \\ \hline
Batch Normalization & -                         & -                 & -             & (32, 18, 18)      & 72                \\ \hline
MaxPooling          & Kernel: $2\times2$        & $2\times2$        & Valid         & (32, 9, 9)        & 0                 \\ \hline
Dropout             & Rate: $0.5 $              & -                 & -             & (32, 9, 9)        & 0                 \\ \specialrule{.2em}{.1em}{.1em}
Flatten             & -                         & -                 & -             & (2592)            & -                 \\ \hline
Fully connected     & Reg: L2 (0.0001)          & -                 & -             & (64)              & 165952            \\
                    & Activation: Softmax       & -                 & -             & -                 & -                 \\ \hline
Fully connected     & Reg: L2 (0.0001)          & -                 & -             & (32)              & 2080              \\
                    & Activation: Softmax       & -                 & -             & -                 & -                 \\ \hline
Fully connected     & Activation: Sigmoid       & -                 & -             & (1)               & 33                \\ \hline
\end{tabular}
\end{minipage}
\end{table*}

\section{Method: a neural network model for merger classification}
\label{sec:deepmerge}

An algorithm that distinguishes between classes of objects uses features that are indicative to those objects to determine key differences.
These features can be and are often clearly defined in terms of physical properties of objects. 
As such, features can be used in algorithms that relate strongly to physical intuition, like the matched filter~\citep{SZ2014,HZ2015}. 
Pre-designated features can also be used in machine learning algorithms, like support vector machines or random forests~\citep{CV1995,HO1995}.
Deep learning algorithms, on the other hand, are optimized during the training phase to identify these features that are primarily responsible for distinguishing between object classes~\citep{LB1998,CB1998}. 

Convolutional neural networks (CNN) are a class of deep learning algorithms specializing in working with images. 
They are usually comprised of three types of layers.
The convolutional layer replaces the simple fully-connected layer. Instead of having a one-dimensional layer of neurons, each having one weight and one bias, convolutional layers have multiple weights and biases, where each weight represents a pixel of a convolutional filter.
This filter is convolved with the input image to produce a two-dimensional representation of the image known as an activation map, which stores the information about the response of the kernel at each spatial position of the image.
The results of the convolutional layer are then passed through a non-linear function, which helps CNN learn and represent almost any complex function which connects input and output values.
Pooling layers perform downsampling along the spatial dimensions of the activation maps. This decreases the required amount of computation and weights, while also helping to reduce over-fitting.  
CNNs also have fully-connected layers, where all neurons in one such layer are connected to all neurons in the preceding and succeeding layers. 
The last fully-connected layer performs the classification. 

Different CNN architectures can be constructed by sequentially adding these layers. 
Complex architectures like Xception~\citep{C2016} can classify merging galaxies on low redshifts $0.02<z<0.06$ with very high precision of 0.97~\citep{AS2018}. 
The Xception architecture has 36 convolutional layers placed into 14 modules. 
It is based on  “depthwise separable convolutions”, which are performed independently for each channel of the image, followed by a $1\times1$ pointwise convolution across all channels~\citep{C2016}. 

We employ a relatively simple sequential model to classify the merger image data. 
The {\it DeepMerge}\footnote{The code used in this paper is available at: https://github.com/deepskies/deepmerge-public} CNN architecture consists of only three convolutional layers. 
The architecture of the {\it DeepMerge} CNN is presented in Table~\ref{table:arch} and 
visualized in Figure~\ref{fig:arch}, where convolutional layers are yellow, pooling layers are red, and fully connected layers are violet\footnote{Figure was created using PlotNeuralNet code~\citep{IQ2018}.}.
The first convolutional layer has eight filters, $5\times5$ in size, the second convolutional layer has 16 filters, $3\times3$ in size, and the third convolution is done with 32 filters, $3\times3$ in size. 
Each convolution is followed by batch normalization and then pooling, which down-samples by a factor of two. 
In all convolutional layers we use a common activation function used today - Rectified Linear Unit (ReLU).  
The last convolutional layer is then flattened to one dimension. It is followed by three fully-connected layers with 64, 32, and one neuron, respectively.
We use the Softmax activation function in the first and second fully-connected layer, because the CNN performed slightly better compared to the use of the ReLU function in these layers. 
The final fully-connected layer employs the Sigmoid activation function because this layer has only one neuron and produces an output between 0 and 1. The {\it DeepMerge} output is taken as a probability of an object being a merger, and we set the threshold to be $0.5$.
Since our problem is a binary classification problem, we choose binary cross-entropy as our loss function. Optimization is performed by using the Adam optimizer~\citep{KB14}.  

Over-fitting of the network model is mitigated by the use of regularization through dropout of $50\%$ during training, applied after all convolutional layers (this is higher than typical dropout rate, but lower rates resulted in quite early over-fitting). 
We also use L2 regularization (also called Ridge Regression) applied on the weights via a kernel regularizer with penalty term $\lambda=0.0001$ in the first two dense layers. 
In case of Ridge Regression, the regularization term is the sum of squares of all the feature weights (multiplied by the penalty term). 
In this case, weights are forced to be small but not zero, which makes L2 a good choice to tackle over-fitting issues.  

We trained the {\it DeepMerge} CNN on both pristine and noisy images. 
In both cases, we only use two HST filters, ACS F814W and WFC3 F160W. 
We initially set our training to last for 500 epoch, but we also include early stopping. 
Early stopping is performed by monitoring the loss function, and training is stopped if validation loss does not drop at all for $50$ epochs.
We use the same architecture and the same set of hyperparameters, on both types of images. Learned weights in case of pristine and noisy images are of course different. The fact that there is a difference, allows us to make an interesting stark comparison between the two data sets. The network performs better on pristine images in comparison to more realistic images, and early stopping enables us to tackle over-fitting. We saved the model with the best weights derived during training (weights which maximize validation accuracy).  

Training and testing our model was done on HP Compaq Elite 8300 CMT, which has Intel Core i5-3470 with 4 cores (3.2GHz), and 16GB of RAM. Training the model for 500 epochs on this machine takes around $18$ hours.

\section{Results}
\label{sec:results}

\begin{figure*}
   \centering
  \includegraphics[width=\linewidth]{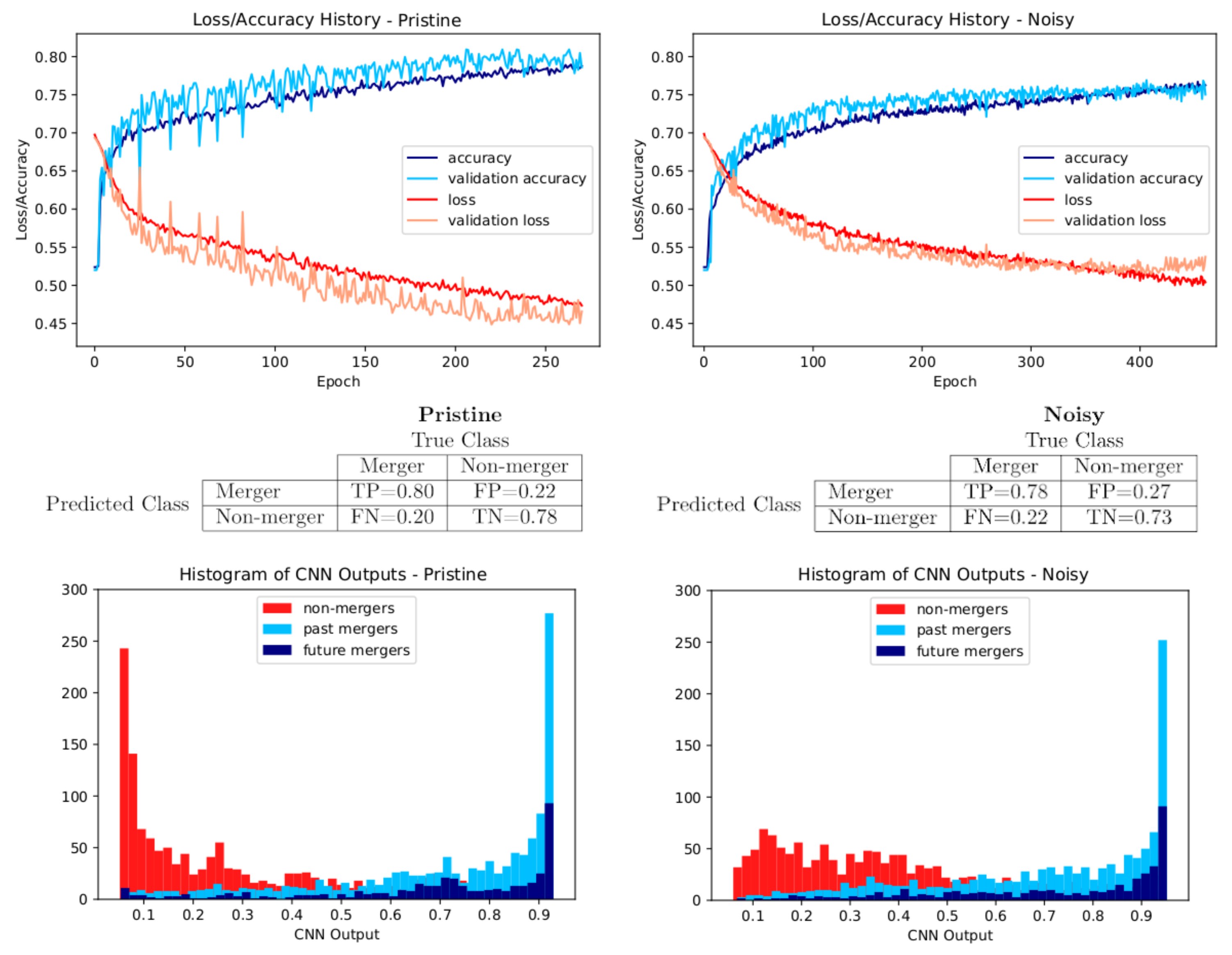}
\caption{
\textbf{Top row:} Accuracy and loss functions and their evolution with training epoch: training on pristine images (left panel) and noisy images (right panel). 
On both panels loss function calculated for running the architecture on training sample of images is presented with red, while loss function after using the validation sample of images is presented in light red. 
Furthermore, training accuracy is plotted using blue line, while validation accuracy is plotted using a light blue line. 
\\
\textbf{Middle row:} Normalized confusion matrices of {\it DeepMerge} CNN, after classifying pristine (left) and noisy (right) test set of images.
\\
\textbf{Bottom row:} Histograms showing the output of {\it DeepMerge} CNN used on the test sample of images, with left panel showing results in case of pristine images, while right panel shows results in case of noisy images. 
Non-mergers are presented in red, future mergers in blue and past mergers in light-blue.
}
\label{fig:all}
\end{figure*}

We present details of the training process and results of the trained models.
We trained the {\it DeepMerge} CNN with early stopping, such that the number of epochs reached $271$ and $461$ for pristine and noisy images, respectively.  
The best model --- deemed by the highest classification accuracy on the validation sample --- was achieved after $227$ and $407$ epochs in the case of pristine and noisy data, respectively.
Overall, the accuracy of classification (on the test set) of the {\it DeepMerge} CNN for pristine and noisy images is $76-79\%$, with pristine images having somewhat higher accuracy.
The test accuracy with pristine images may be attributed to the fact that there is no noise to obscure important discriminating features.

We present the performance results through a set of conventional metrics --- the histories of loss and accuracy during training and validation, the confusion matrix, distributions of CNN probabilities for mergers, non-mergers, and past mergers, the receiver-operator characteristics (ROC) curve, and the area under the curve (AUC).
Mergers and non-mergers correctly classified are true positives (TP) and true negatives (TN), respectively. 
Incorrectly classified mergers and non-mergers are false negatives (FN) and false positives (FP), respectively.
The confusion matrix summarizes classification success through counts or fractions of TP, TN, FP, and FN.
The ROC curve graphically shows the trade-off between Sensitivity (TP/(TP+FN)) and Specificity (TN/(TN+FP)) --- i.e. trade-off between true-positive rate and false-positive rate. 
The AUC summarizes the ROC curve: for example, where the AUC is close to unity, classification is successful, while an AUC of 0.5 indicates the model performs as well as a random guess.

The top row of Figure~\ref{fig:all} shows the accuracy and loss history during training and validation for pristine (left) and noisy images (right). 
The training for the model of noisy images require almost twice as many epochs to achieve the best validation accuracy. 
We present the normalized confusion matrices for our test sample of pristine (left) and noisy (right) images in the middle row of Figure~\ref{fig:all}.
Each field in the confusion matrix shows the percentage of merger images classified as TP and FN, as well as non-merger images classified as TN and FP. 

\begin{figure*}
   \centering
  \includegraphics[width=\linewidth]{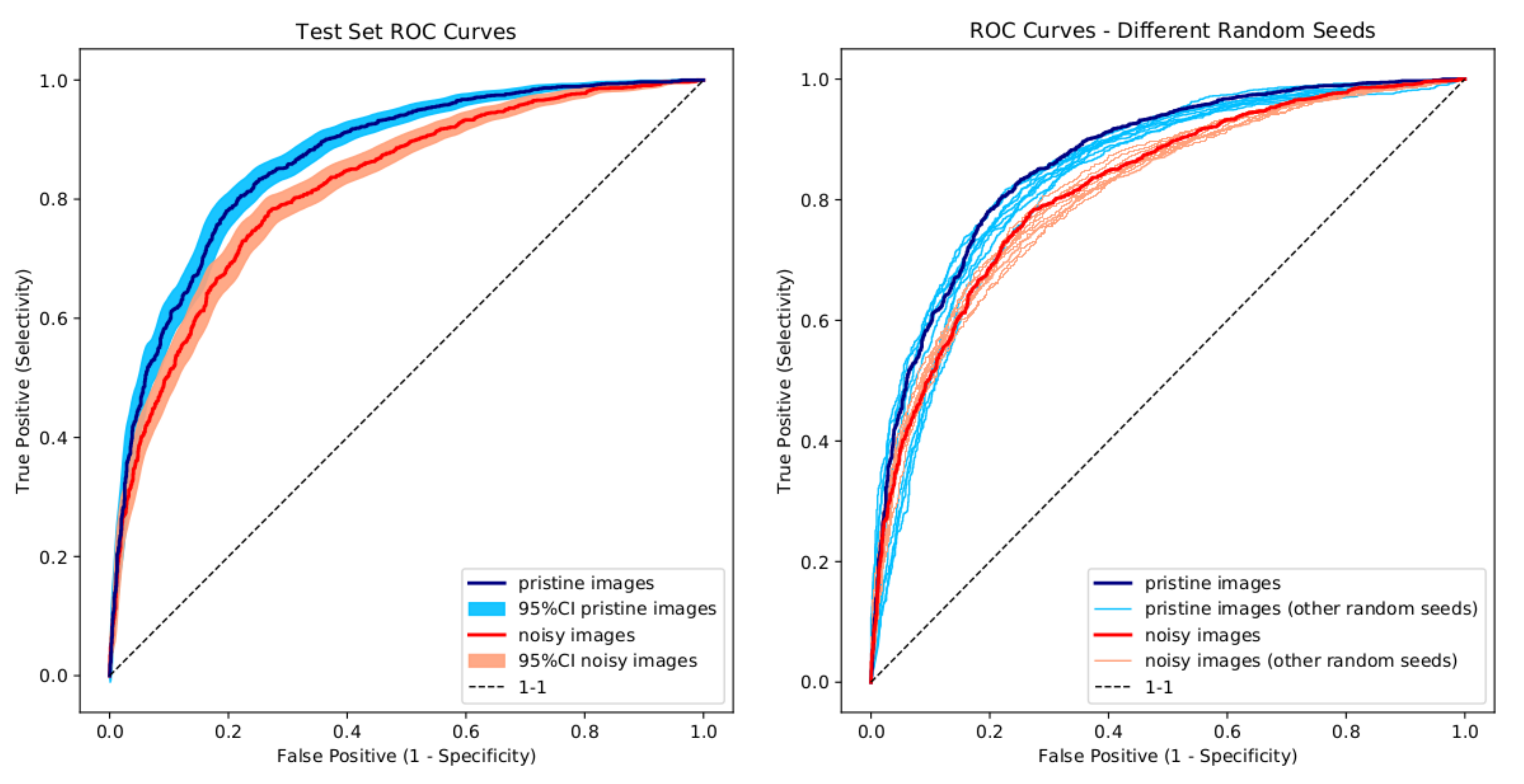}
\caption{
ROC curves of the {\it DeepMerge} classifier, after training with pristine images (blue), and noisy images (red). The results show the classification performance of the model with the best weights, applied to the test sample of images. 
On the left panel we plot $95\%$CI bands in the vertical direction (for true positives) generated by bootstrapping (pristine images - light blue band, noisy images - light red band). 
The right panel shows the same pristine (blue) and noisy (red) ROC curves, compared to test set ROC curves derived when different random seeds were used to separate images into train, test and validation samples. 
In case of pristine images these ROC curves are plotted with light-blue lines and in case of noisy images with light-red lines.}
\label{fig:roc}
\end{figure*}

Figure~\ref{fig:roc} (left panel) presents ROC curves for classification performed on the test set --- the pristine data is in blue (AUC=0.86) and the noisy data is in red (AUC=0.82). 
Error bands on the figure represent $95\%$ confidence intervals ($95\%$ CI) in the true positive, generated by bootstrapping 1000 samples with replacement.

\begin{figure}
\centering
  \includegraphics[width=0.9\linewidth]{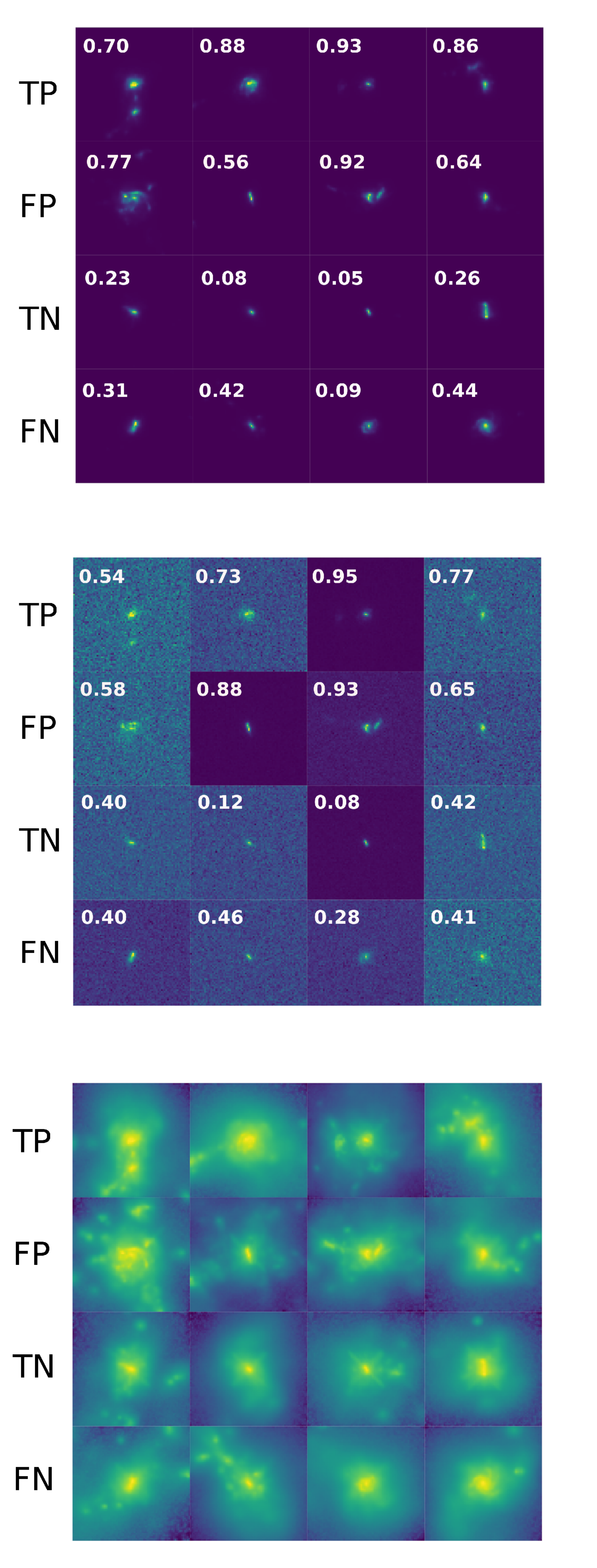}
\caption{Examples of TP, FP, TN and FN. 
Top panel shows examples drawn from pristine test images. 
Middle panel shows the same images but from our noisy test sample. 
Same pristine images, but drawn with logarithmic colormap normalization, are presented on the bottom panel. 
Top and middle panel also include the output value of our CNN, which is used to classify objects (non-mergers have output bellow $0.5$, while mergers are above this value).} \label{fig:examples}
\end{figure}

Next, in Figure~\ref{fig:examples}, we show examples of images from the test set in the top and middle panels for the pristine and noisy images, respectively. 
In each panel of images, the rows --- from top to bottom --- show TP, FP, TN, and FN examples, respectively. 
Overlaid are the output values of network for each image.
In the bottom panel of the same figure, we plot the same pristine images, but with a logarithmic color-map normalization to better show the structure of these objects.
Since the top and bottom panels show the same images, these output values can show how training and testing with pristine and noisy images changes the output result for the same chosen examples.

The performance of a classifier can also be described by the Precision (``purity'' or ``positive predictive value''; TP/(TP+FP)), Recall (``completeness'' or ``true positive rate''; TP/(TP+FN)) and $\mathrm{F1\,Score} = 2\,\frac{\mathrm{Precision} \times \mathrm{Recall}}{\mathrm{Precision} + \mathrm{Recall}}$.
This metric can sometimes be even more indicative of a classifier performance in comparison to accuracy (for example in cases where one class is much more populated). The {\it DeepMerge} CNN trained on pristine images has precision of $0.81$, and recall of $0.80$.
When training with noisy images {\it DeepMerge} CNN has precision of $0.77$, and recall of $0.78$. 

In the case of balanced samples, a useful scoring method is the Brier score (BS). 
It represents the mean squared error (MSE) between predicted probabilities (between 0 and 1) and the expected values (0 or 1), and hence can be thought of as a measure of the "calibration" of a set of probabilistic predictions. For instance, if a binary classifier is well calibrated, out of all samples classified as positive class with output probability of 0.9, approximately $90\%$ should actually belong to the positive class.
Finally, Brier score summarizes the magnitude of the forecasting error and takes a value between 0 and 1 (with better models having BS close to 0).
The Brier score for our \emph{DeepMerge} classifier is 0.15 for pristine images, and 0.17 for noisy images.

~\cite{SR2018} train a RF classifier on the same sample of galaxies from Illustris simulation (but they use galaxies with $0.5<z<4$). 
They show performance of the RF classifier for every redshift they used.
In the case of redshift $z=2$ (which we used in this paper), and using a balanced samples of mergers and non-mergers their precision and recall are both $\approx 0.7$ (their Figure 15). 
The authors show that the RF classifier has superior performance compared to one or two-dimensional statistics that are commonly used to classify mergers. 
Based on the CNN performance, we show that {\it DeepMerge} CNN outperforms the RF classifier. 

\begin{table*}[t]
   \centering
   \noindent\begin{minipage}[b]{0.99\textwidth}
   \centering
    \caption{Performance metrics of the {\it DeepMerge} CNN. 
    The table shows Area Under the Curve (AUC), Accuracy, Precision (purity, positive predictive value), Recall (completeness, true positive rate or sensitivity), F1 score and Brier score for our test set of images. 
    Errors in the table represent $95\%$CI generated by bootstrapping. First two columns show results when CNN is both trained and tested with pristine and with noisy images, respectively. Second two columns show results when trained on pristine / tested on noisy images, and trained on noisy / tested on pristine images, respectively.}
  \label{table:perf}
  \centering
  \begin{tabular}{|l | c c| c c|}
\hline
 
\multirow{2}{*}{\diagbox[width=65pt]{Metric}{Train\\Test}} & &  & &  \\
                              & Pristine  & Noisy  & Pristine & Noisy \\ 
                             &  Pristine & Noisy & Noisy & Pristine \\\hline
AUC          & $0.86\pm 0.01$    &  $0.82\pm 0.01$     &  $0.53\pm 0.02$      &   $0.79\pm 0.02$       \\
Accuracy     & $0.79\pm 0.01$    &  $0.76\pm 0.01$     &  $0.47\pm 0.02$      &   $0.56\pm 0.02$       \\ 
Precision    & $0.81\pm 0.02$    &  $0.77\pm 0.02$     &  $0.74\pm 0.01$      &   $0.55\pm 0.02$     \\
Recall       & $0.80\pm 0.02$    &  $0.78\pm 0.02$     &  $0.03\pm 0.009$      &   $0.98\pm 0.007$     \\
F1 score     & $0.81\pm 0.02$    &  $0.77\pm 0.02$     &  $0.06\pm 0.02$      &   $0.71\pm 0.01$     \\
Brier score & $0.15\pm 0.007$    &  $0.17\pm 0.007$     &  $0.42\pm 0.01$      &   $0.30\pm 0.01$       \\\hline
\end{tabular}
\end{minipage}
\end{table*}

\section{Discussion}
\label{sec:discussion}
We present a discussion, in which we compare the {\it DeepMerge} network model to other models in the literature, perform a variety of experiments to explore its sensitivity to training data, and probe interpretability of its predictions. 

\subsection{Comparison to other CNN architectures}
A similar galaxy merger classification was performed with CNNs in~\cite{PW2019}. 
In one scenario, the authors train their network with real SDSS observational image data~\citep{DK10a,DK10b}, in the redshift range $0.005<z<0.1$, to achieve very high classification accuracy of $91.5\%$. 
In another scenario, the training set comprises EAGLE simulation~\citep{MA16}, where simulated images are processed to mimic SDSS observations in the same redshift range. 
It achieved $65.2\%$, $64.4\%$, and $67.4\%$ accuracy in the cases where galaxies are deemed mergers when they are within $100\,\mathrm{Myr}$, $200\,\mathrm{Myr}$, and $300\,\mathrm{Myr}$ of the merger event, respectively. 
The last two cases can be compared to our study, because we use the same images as in~\cite{SR2018}, where mergers were selected to be within $250\,\mathrm{Myr}$ from the merger event. 
With these two larger time windows around the merger event,~\cite{PW2019} have precision $0.67-0.68$ and recall $0.56-0.65$, which are lower than the results of the {\it DeepMerge} CNN. 
Table~\ref{table:perf} (two left columns) provides a summary of the performance of the {\it DeepMerge} CNN trained and tested on pristine and noisy images. 
Errors in the table are generated by 1000 bootstrap re-samples (with replacement), and they represent $95\%$CI.

\subsection{Sensitivity to data arrangement}
We performed a test to study the stability of the network training under changes in image data order. 
We consider this to be an important standard diagnostic for any network training to guard against biases in network predictions.
This was done by fixing the random seed before shuffling images prior to their division into training, testing, and validation samples. We ran 10 different random seed experiments for both pristine and noisy sample.
On the right panel of Figure~\ref{fig:roc}, we show the ROC curves for all the experiments with the random seeds, performed on the test sample, including the best-performing network (pristine -- blue line and noisy -- red line), for pristine (light-blue lines) and noisy images (light-red lines).
In general, ROC curves vary up to 20\% in the TP rate below FP rates of 20\%. 
In Table~\ref{table:seeds} we give the intervals in which test set AUC, accuracy, precision, recall, F1 score and Brier score are located, for both pristine and noisy case, when different random seeds are used for shuffling images. 
The AUC is in the range $0.83-0.87$ and $0.81-0.83$ in case of pristine and noisy images, respectively. 
The accuracy, F1 score, and Bier score exhibit behavior similar to the AUC, and precision and recall have slightly larger intervals.
In case of precision this is caused by few runs with lower TN rates (below 0.7), which makes FP rate larger and in turn lowers precision. 
Recall interval is, on the other hand, affected by few runs which have slightly lower/higher TP rate than the others.

\begin{table*}[t]
   \centering
   \noindent\begin{minipage}[b]{0.99\textwidth}
   \centering
    \caption{The intervals in which the test set classification scores (Area Under the Curve -- AUC, Accuracy, Precision, Recall, F1 score and Brier score) are located when different random seeds are used to shuffle pristine and noisy images before they are placed into training, testing and validation samples.}
  \label{table:seeds}
  \centering
  \begin{tabular}{|l | c c|}
\hline
  
\multirow{2}{*}{\diagbox[width=65pt]{Metric}{Train\\Test}} & \multicolumn{2}{c|}{}\\ 
& Pristine & Noisy \\
& Pristine & Noisy \\\hline
AUC              & $0.83-0.87$    &  $0.81-0.83$    \\
Accuracy         & $0.76-0.79$    &  $0.73-0.76$   \\ 
Precision        & $0.72-0.81$    &  $0.73-0.80$ \\
Recall           & $0.76-0.88$    &  $0.70-0.78$   \\
F1 score         & $0.78-0.81$    &  $0.74-0.77$ \\
Brier score      & $0.15-0.18$    &  $0.17-0.18$    \\\hline
\end{tabular}
\end{minipage}
\end{table*}

\subsection{Sensitivity Tests: noise}
\label{sec:noise}
Next, we test network efficacy and sensitivity when presented with image types that it was not trained on --- i.e., we classify pristine images using CNN trained on noisy images and vice versa. 
In this type of situation, performance should be worse compared to CNN both trained and tested on the same type of images, but some classification might still be possible. 
The network trained on pristine images is incapable of classifying noisy images and assigns most of the images to non-merger class (AUC=$0.53$). 
When trained on pristine images, the network can likely learn subtler characteristics more easily, which increases accuracy when classifying the pristine test set, but also makes it unusable for noisy test set in which detailed structures are more likely to be obscured.

The network trained on noisy images can classify pristine images somewhat better -- for the random seed and parameter choices presented in detail in this paper, the CNN has AUC=$0.79$. 
In this case many more images are assigned to the merger class, and the accuracy of classification is only $56\%$.
In this type of tests (with other random seeds) we generally noticed somewhat better performance in CNNs trained on noisy images. 
The reason for this could be that the noise added to the pictures is helping {\it DeepMerge} CNN see the big picture and classify mergers without focusing on smaller-scale details that are more visible in pristine images (filaments, substructures, very faint halos etc.), which introduce more diversity of structure --- making classification more difficult. 
For this reason the CNN trained on noisy images can probably generalize better and classify some pristine images. 

The performance of the {\it DeepMerge} classifier in both cases where training and testing was done on different types of images is also given in Table~\ref{table:perf} (columns three and four). 
Although these CNNs never performed as good as the architecture trained and tested on the same type of images, one particular version of CNN trained on noisy images, classified pristine images with fairly high test accuracy of $74\%$ (TP=0.87, TN=0.60) and had AUC=0.83.

\subsection{Merger sub-groups}

We tested how the performance of the {\it DeepMerge} CNN classification changes within two merger subgroups. In this paper we follow~\cite{SR2018}, who define mergers as all objects which are withing $250\,\mathrm{Myr}$ from the merger event. 
We split our sample of mergers into past mergers (mergers completed within the past $250\,\mathrm{Myr}$ of the present snapshot) and future mergers (mergers that will take place within the $250\,\mathrm{Myr}$ after the present snapshot), 
In Figure~\ref{fig:all} (bottom row), we present distributions of the classification results for these different merger subgroups when tested on pristine images (left) and noisy images (right). 
Non-mergers are presented in red, future mergers in blue, and past mergers in light-blue. 
In both merger subgroups (past and future) and for both pristine and noisy images, most results are close to $1$.
The CNN is only slightly less certain when classifying noisy non-mergers, with more values further away from zero, but even in this case most non-mergers are still classified correctly. 

For galaxies in our sample for which we have concentration and $\mathrm{M}_{20}$ available (see Table 2 from~\cite{SR2018}), we tested whether the output probabilities were influenced by these parameters, but we found no connection. 
This appears to differ from the results of~\citet{SR2018}, who find that morphological parameters indicating the presence of a bulge have high importance for past mergers in the RF classifications. More precise conclusions in case of our CNN classification might be possible if these parameters were available for all galaxies in our sample. 

\begin{figure*}
\centering
  \includegraphics[width=\linewidth]{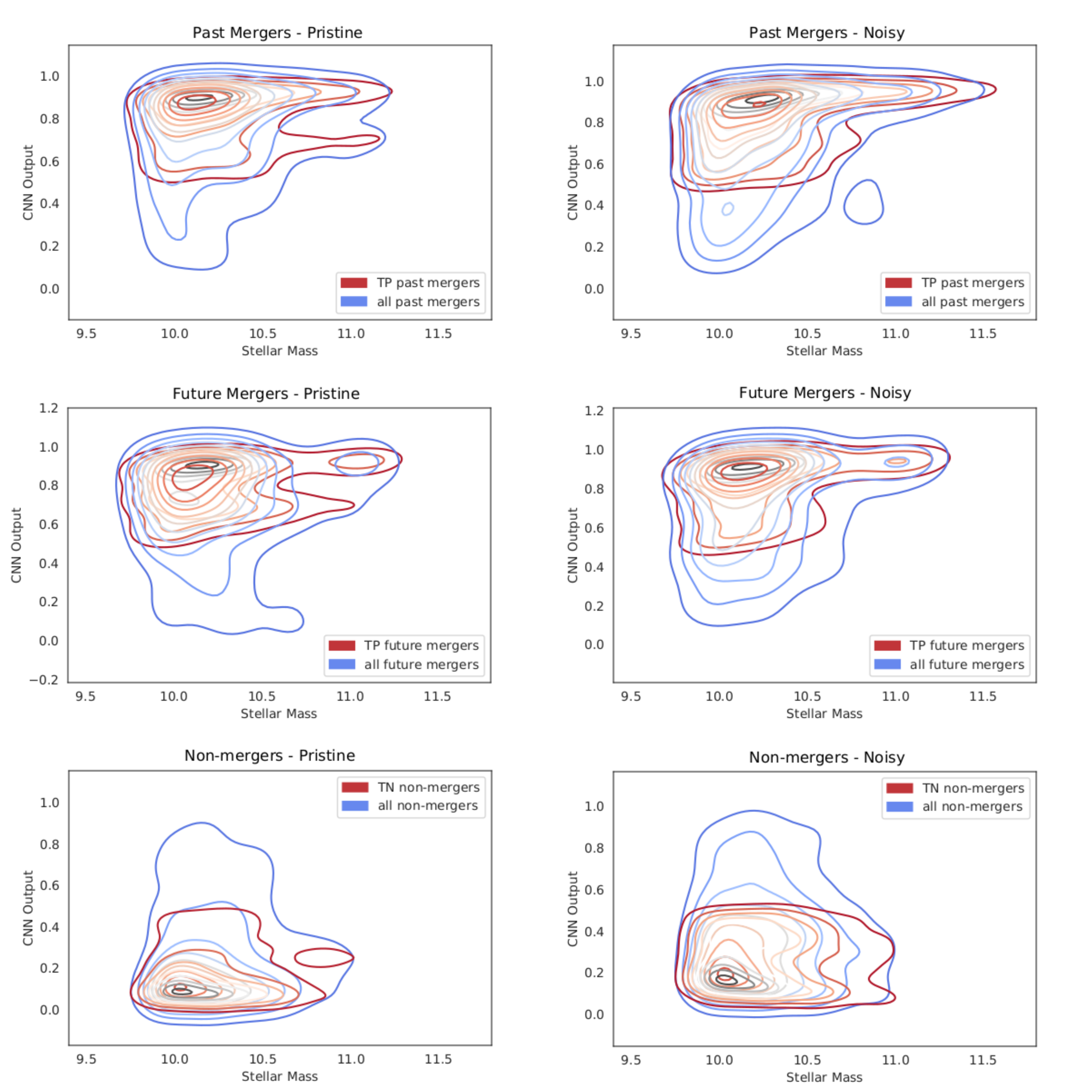}
\caption{Histograms of the distribution of output probabilities and galaxy stellar masses $\log_{10}{M_{*}/M_\odot}$, for past mergers, future mergers and non-mergers (from top to bottom, respectively). Histograms of the entire classes are plotted in blue, while TPs (for past and future mergers) and TNs (for non-mergers) are plotted in red. Pristine and noisy case are plotted in left and right column, respectively.} \label{fig:mass}
\end{figure*}

We also examine the impact of classification on selecting for different physical aspects of merger populations --- in particular, stellar mass. 
We find that there is no significant bias in stellar mass during classification of mergers. 
This is illustrated in Figure~\ref{fig:mass}, which shows 2D histograms of the  distribution of the output probabilities against the stellar mass $M_{*}$, given as $\log_{10}{M_{*}/M_\odot}$, where $M_\odot$ is the solar mass. Panels on the left show results for our pristine test set, and panels on the right for the noisy test set. Past merger, future merger and non-merger histograms are plotted from top to bottom, respectively. On all histograms we plot all past and future mergers and non-mergers from the test sample with blue lines, while TPs in case of mergers and TNs in case of non-mergers we plot in red. Both mergers and non-mergers in our sample have very similar stellar mass distributions, with most objects having $\log_{10}{M_{*}/M_\odot}$ between $9.75-10.5$. Figure~\ref{fig:mass} shows that most of incorrectly classified mergers and non-mergers are lower stellar mass objects.

\subsection{Interpretability of CNN Predictions}

Finally, we seek to interpret the neural networks and identify the features deemed by the neural network to be important in distinguishing mergers from non-mergers.
One technique is the ``saliency map'', first developed by~\cite{SV2013}, which can be produced by computing the gradient of the CNN output values with respect to the input image. 
This gradient can be used to describe how the CNN output changes with respect to a small changes in any of the pixels of the input image. 
For example, in~\cite{PB2019}, saliency maps are used to show that ridge-like features are key for their CNN models to distinguish between different levels of magnetization in turbulence simulations.

A more recent technique, Gradient-weighted Class Activation Mapping  ~\citep[Grad-CAM;][]{SC16} produces a localization map in which the most important regions for classification are highlighted. 
Grad-CAM calculates class-specific gradients $\frac{\partial y^c}{\partial A_{ij}^k}$ of the output score $y^c$ (score for class $c$) with respect to the activation maps (i.e. feature maps) of the last convolutional layer $A_{ij}^k$ (dimension of the feature map is $i\times j = Z$ pixels, and $k$ lists all feature maps of the last convolutional layer). 
These gradients are global-average-pooled to calculate the importance weights $\alpha_k^c $:
\begin{equation}
\alpha_k^c = \frac{1}{Z}\sum_{i}\sum_{j}\frac{\partial y^c}{\partial A_{ij}^k}.
\end{equation}
Grad-CAM maps are then produced from weighted combination of feature maps, followed by a ReLU function (which extracts all output positive regions for the class we are interested in):
\begin{equation}
L^c_\mathrm{Grad-CAM} = \mathrm{ReLU}\left(\sum_k \alpha^c_k A^k\right).
\end{equation}
We produce a coarse localization map in which the most important regions for classification are highlighted.
With this technique, we use the spatial information contained in the feature maps of the final convolutional layer, which would get completely lost in the later dense layers.

\begin{figure*}
\centering
  \includegraphics[width=0.90\linewidth]{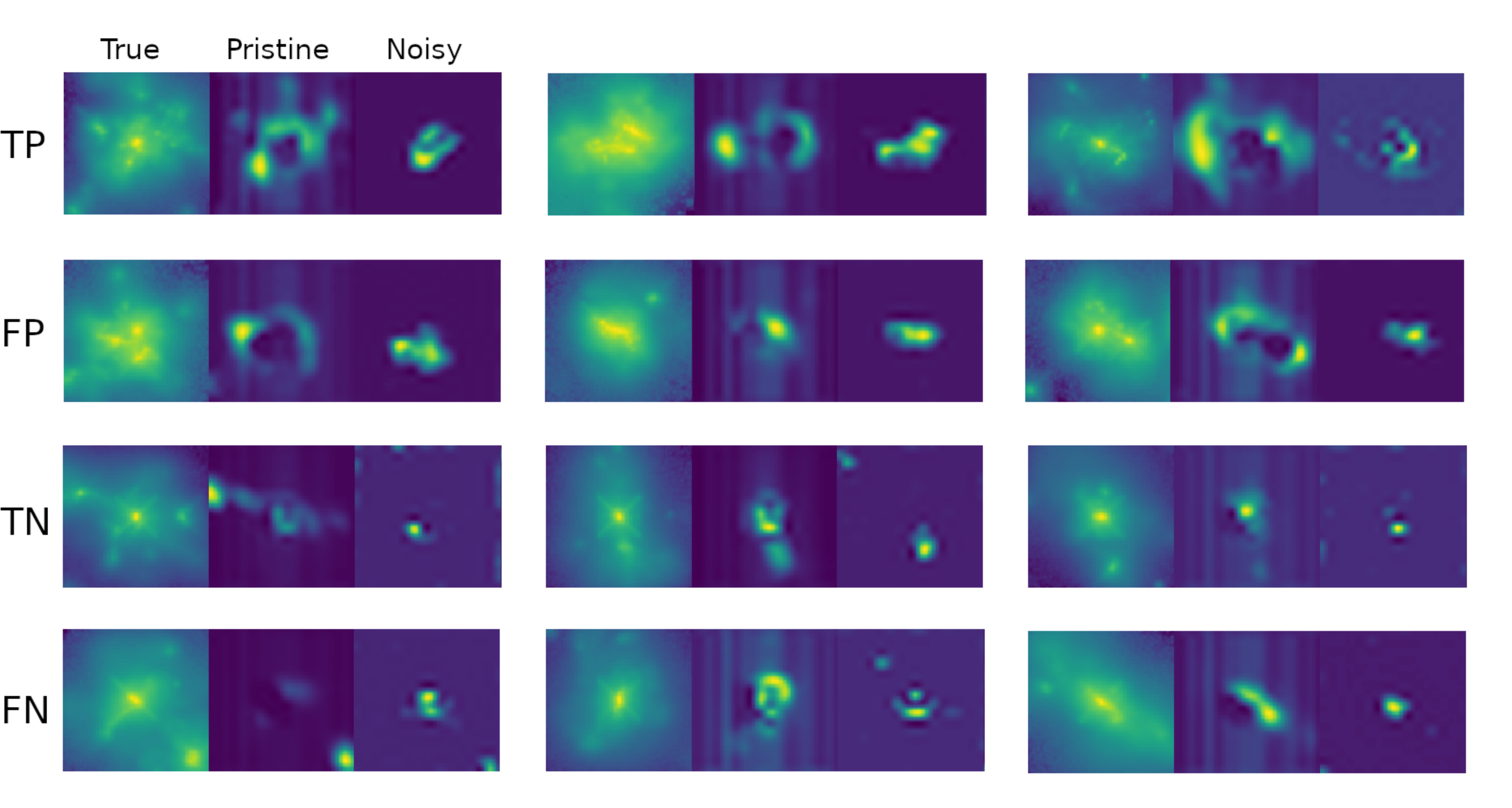}
\caption{
Gradient-weighted Class Activation Maps (Grad-CAMs) highlight the most important regions that the {\it DeepMerge} CNN uses to classify images. 
We choose images that were classified with high certainty in both pristine and noisy cases to show the difference between the important regions and the influence of the added noise. 
Rows from top to bottom show examples of images classified as TP, FP, TN and FN, respectively. For each group we give three different examples. We plot the galaxy image on the left (with logarithmic colormap normalization, to make faint details more visible), Grad-CAM from the pristine image case in the middle, and Grad-CAM from the noisy image case on the right.
\label{fig:grad_CAM}
} 
\end{figure*}

\begin{figure*}
\centering
  \includegraphics[width=0.8\linewidth]{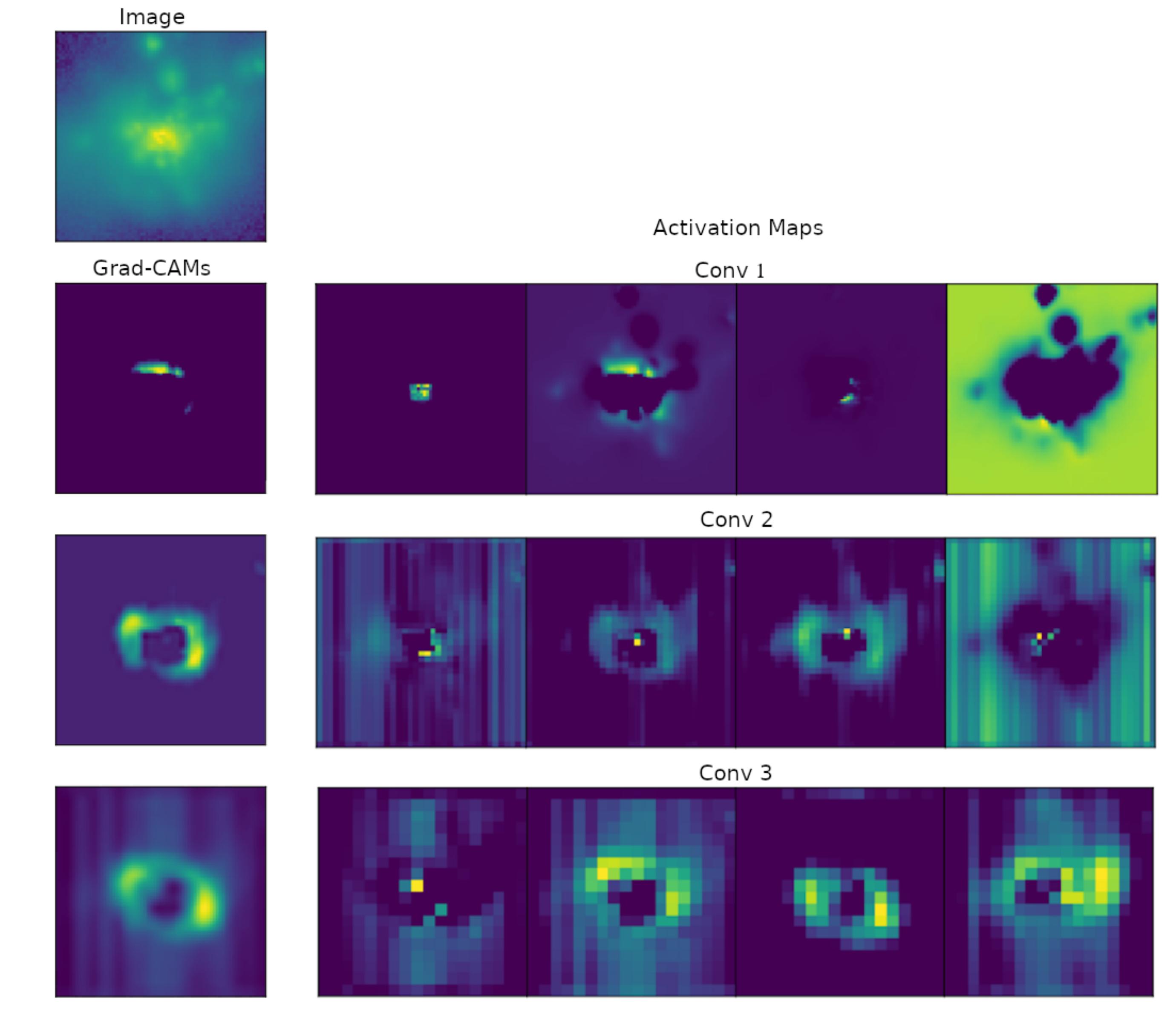}
\caption{Grad-CAM localization maps (first column on the left) and activation maps for four randomly chosen filters (all other columns on the right), for an example pristine image (plotted on top of the first column on the left). 
Rows from top to bottom of first column on the left show Grad-CAM maps produced by using first, second and third convolutional layer. 
Activation maps from the first, second and third convolutional layer are also plotted (on the right) in the first, second and third row, respectively.} \label{fig:grad_CAM_comparison}
\end{figure*}

In Figure~\ref{fig:grad_CAM}, we present examples of localization maps in the case of pristine images and noisy images. 
The first and second row show examples of TPs and FPs (all classified with very high probability), and the third and fourth rows show TN and FN examples (all classified with very low probability). 
By plotting Grad-CAMs for the same images with and without noise we can see how the region which CNN finds important changes when noise is added. For all examples we plot the galaxy images (with logarithmic colormap normalization, for more details to be apparent) on the left, Grad-CAM from the pristine case in the middle and Grad-CAM from the noisy case on the right.

In the case of pristine images, these localization maps show that fainter substructures indeed play an important role when an image is classified as a merger. 
In the case of mergers, the CNN seems to look at larger, more complex regions at the periphery of galaxies. 
On the other hand, important regions in case of non-mergers are somewhat smaller and compact. 
As expected, in the the case of noisy images, the CNN does not see fainter structures as well, so objects classified as mergers have a more compact regions which are important, but these regions can still have asymmetric shapes. 
Non-mergers (TNs and FNs) have, on the other hand, very compact important regions. 
In both pristine and noisy cases, all images with output values around 0.5, no matter which class they were classified as, have the size and shape of the most important regions somewhere in between the high-probability classifications, we have presented in Figure~\ref{fig:grad_CAM}.

When using Grad-CAM to visualize the important regions of the image, the convolutional layer used should be close to the layer whose outputs we want to visualize. 
To show how Grad-CAM localization maps degrade with distance from the convolutional layer to the output layer in Figure~\ref{fig:grad_CAM_comparison} we also show Grad-CAM maps for the classification of one example pristine image (plotted on top of the first column of images). 
In the first column, we plot Grad-CAM results using the first, second and third convolutional layer (from top to bottom). 
As you can see, the quality of the localization increases from top to bottom, as the convolutional layer used becomes closer to the output layer. 
It is also interesting to compare the localization map produced by Grad-CAM with the activation maps of that convolutional layer, because the information contained in these maps combined with the gradients of the outputs is what produces Grad-CAM maps.
On the right side of the Figure~\ref{fig:grad_CAM_comparison} we plot activation maps for randomly chosen filters from the first, second, and third convolutional layers (from top to bottom), which have 8, 16 and 32 filters in total, respectively.

\subsection{Domain transfer and working with real astronomical images}

In this paper we show that deep learning can be a very useful method for classification of simulated high-redshift merging galaxies. 
With the future launch of large telescopes like WFIRST, large high-redshift observational data sets will become available. 
This will open the door for the application of deep learning models for unlabeled observed images. 
The simulated data for training neural networks must closely mimic the observational data.
However, simulated images may only asymptotically approach absolute realism. 
Discrepancies between simulated and observational data are likely to persist due to a number of factors: 
approximations used in physical modeling due to incomplete knowledge of the physical system;
approximations used to reduce the computational demand;
uncertainties introduced by imperfect modeling of the telescope and the night sky and Earth's atmosphere in case of Earth based telescopes. 
This weakness of simple deep learning algorithms was in part demonstrated in \S\ref{sec:noise}, where we show that the performance of the {\it DeepMerge} model drops when the network that is trained on pristine images assesses noisy images (and vice versa).

There are a variety of approaches for addressing discrepancies when working with data from different domains (for example simulated and real data).
Domain adaptation methods build mappings between the source and the target domains so that the classifier learned for the source domain can also be applied to the target domain~\citep{ZQ2019,ZZ2020}. 
Other approaches, like Domain Adversarial Networks~\citep{GU2015}, include finding a domain-invariant latent feature space.
This type of classifier would only use features present in both domains, which would allow for classification of real unlabeled observations without a loss in accuracy. 
In our follow-up work, we will use domain transfer methods to improve {\it DeepMerge} classification and allow a domain shift between pristine and noisy data sets. 
The same methods will also be applied to shifting from our simulated to real images. 
This will allow us to build a well-performing classifier based on simulated images that will also have the capability of classifying real images with high certainty.

\section{Conclusion and Outlook}
\label{sec:conclusion}

The study of distant galaxy mergers during the period of cosmic high noon presents an opportunity to study the time where most stellar mass was assembled, critical for understanding galaxy evolution.

In this work, we demonstrate the use of a simple neural network to identify high-redshift ($z=2$) merging events with state-of-the art accuracy.
We distinguish between mergers and non-mergers by training a deep neural network ({\it DeepMerge}) that has three convolutional layers and three fully-connected layers. 
We develop networks both for pristine images and those with observational noise that mimics HST. 
We also show that {\it DeepMerge} CNN outperforms the random forest classifier from ~\cite{SR2018} on the same simulated data from the Illustris-1 simulation~\citep{VG2014a,VG2014b}. 
Previous studies of galaxy mergers using CNNs used images of galaxies at much lower redshifts of $z<0.1$ ~\citep{AS2018,PW2019}, and they showed that CNNs can be a very good tool for merging galaxies classification.

We performed a number of experiments to explore the sensitivity of the neural network to data set order and image quality. 
We also analyzed the selection function for mergers in the context of stellar mass and merger class.
Finally, we explore Grad-CAM method to interpret the neural network sensitivities and determine which features it deemed useful for distinguishing merging events.

Future work includes applying this network technique to additional redshift ranges and to real-sky data, and to pursue a hybridization with morphological feature-based modeling.
With larger data sets, it will also be important to test more complex network architectures.
This work may also lend itself to discriminating between merging systems and projected systems and the much-anticipated deblending problem for large, deep cosmic surveys. 
Moreover, there is a positive outlook for predicting physical parameters of merging galaxies and in doing so, learning more about galaxy mergers.
Finally, this works takes another significant step toward the classification of the full range of astronomical objects.

\section*{Acknowledgments}

\subsection*{Author Contributions}

A.~\'Ciprijanovi\'c performed all the neural network tests and development, as well as the data analysis, and scientific direction.

G.~Snyder provided knowledge and consultation on merging of galaxies, as well as the data sets, and provided scientific direction.

B.~Nord contributed to CNN architecture design, the analysis of science product, provided scientific direction, guidance on analysis neural network output and project management.

J.E.G.~Peek provided the initial problem formulation, guidance on saliency evaluation, and feedback/edits on draft.

\begin{figure*}
\centering
  \includegraphics[width=0.90\linewidth]{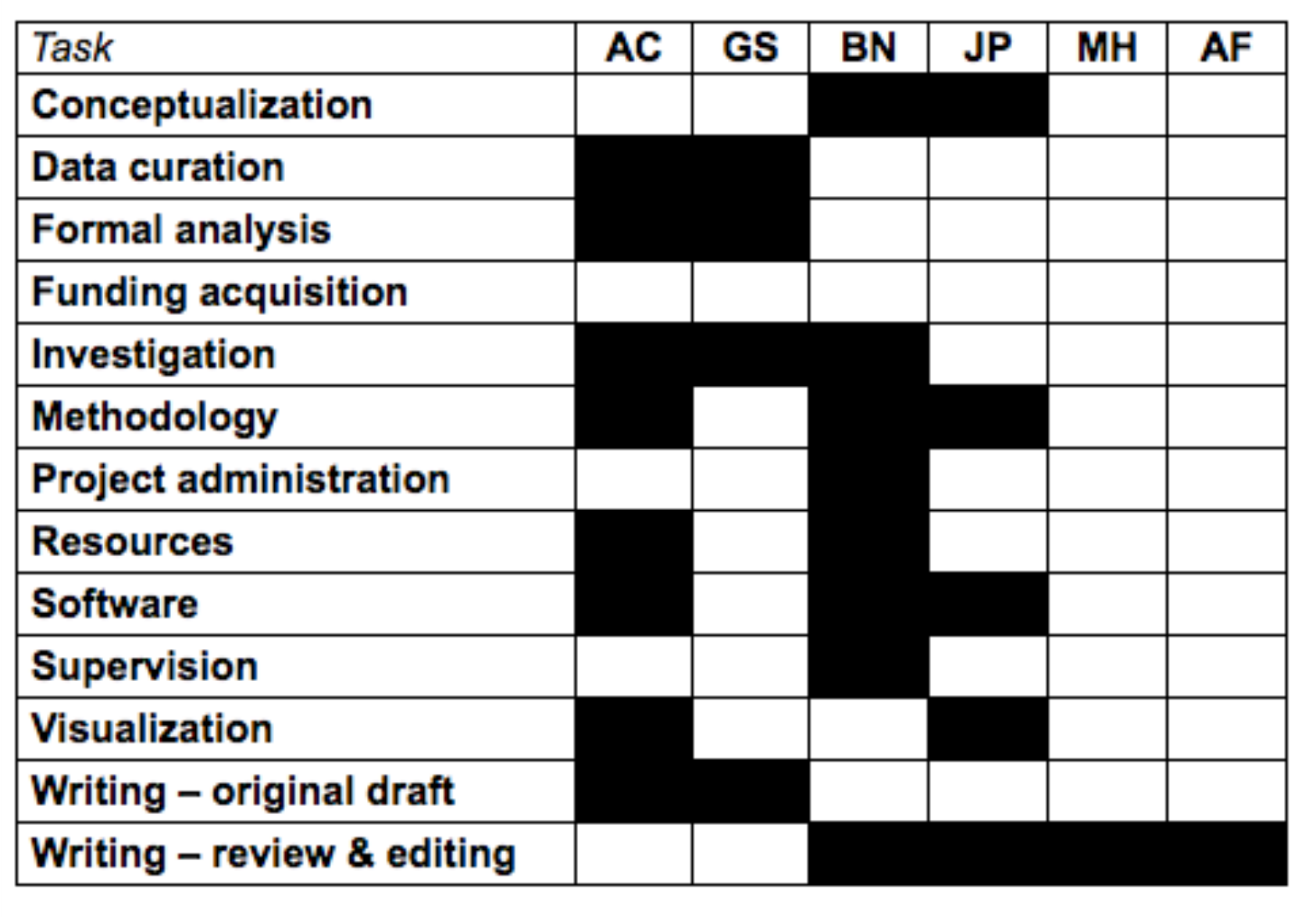}
\caption{Matrix of contributions from authors for quick reference.
\label{fig:contribution_matrix}
} 
\end{figure*}

We present a summary of these contributions in the Contribution Matrix in Fig~\ref{fig:contribution_matrix}.
\newline
This work is supported by the Deep Skies Community (\url{deepskieslab.com}), which helped to bring together the authors and reviewers. 
We thank M. Haas and A. Farahi for valuable insights and comments. We also thank the anonymous referee for their comments, which helped improve this paper.
The authors of this paper have committed themselves to performing this work in an equitable, inclusive, and just environment, and we hold ourselves accountable, believing that the best science is contingent on a good research environment.

The work of A. \'Ciprijanovi\'c is supported by the Ministry of Science of the Republic of Serbia under project number 176005.

This manuscript has been authored by Fermi Research Alliance, LLC under Contract No. DE-AC02-07CH11359 with the U.S. Department of Energy, Office of Science, Office of High Energy Physics.

The work of G. Snyder and the creation of the simulated image dataset was supported by an HST AR-Theory grant, program number 13887, awarded by the Space Telescope Science Institute, which is operated by the Association of Universities for Research in Astronomy, Inc., under NASA contact NAS 5-26555.

\bibliographystyle{model2-names}
\bibliography{DeepMerge_v2}

\end{document}